\documentclass[aps,prl,twocolumn,floatfix,nolongbibliography]{revtex4-2}

\usepackage{MnSymbol}
\usepackage{amsmath,graphicx,color}
\usepackage{natbib}
\usepackage{placeins}
\setcitestyle{square}
\usepackage{braket}
\usepackage{makerobust}
\usepackage{xcolor}
\usepackage{hyperref}
\hypersetup{%
 colorlinks,
 breaklinks=true,
 plainpages=false,%
 citecolor=blue,
 linkcolor=blue,
 urlcolor=blue,
 bookmarksopen=true,%
 bookmarksnumbered=false,%
 bookmarksdepth=5%
}

\newcommand{\makeauthor}[2]{\newcommand{#1}[1]{{%
 \sffamily\color{#2}{%
 \bfseries\begingroup\escapechar=-1\edef\x{\endgroup\string#1}\x:%
 } ##1}}%
 \MakeRobustCommand#1}
\makeauthor{\rew}{blue}
\makeauthor{\jb}{purple}
\makeauthor{\nk}{red}
\begin{document}

\title[Article Title]{
Interaction-mediated Co-existence of Altermagnetism and Topology
}

\author{Jasmin Bedow, Nitin Kaushal and Marcel Franz}
\affiliation{Department of Physics and Astronomy and Quantum Matter Institute, University of British Columbia, Vancouver, British Columbia BC V6T 1Z4, Canada}

\date{\today}

\begin{abstract}
We study an extended Hubbard model on Lieb lattice at electron filling 2 and 4, and demonstrate that the interactions lead to the simultaneous emergence of altermagnetic order and a topological gap. Using unrestricted Hartree-Fock theory, we evaluate both the altermagnetic and topological order parameters and show that they co-exist with an associated quantum spin Hall effect in a large region of the phase diagram, which we further support using exact diagonalization. Moreover, we demonstrate that inversion-symmetry breaking terms can also mediate second-order topological phases with associated corner modes.
\end{abstract}

\maketitle

%%%%%%%%%%%%%%%%%%%%%%%%%%%%%%%%%%%%%%%%%%%%%%%%%%%%%%%%%%
%%%%%%%%%%%%%%%%%%%%%%%%%%%%%%%%%%%%%%%%%%%%%%%%%%%%%%%%%% }

{\it Introduction.~} 
Coulomb repulsion can both generate magnetically ordered states \cite{Hirsch1985,Kaushal2025} and open a topologically non-trivial insulating gap with associated protected boundary modes \cite{Raghu2008}. Whether these two effects can be generated simultaneously is far less clear.
To answer this question, altermagnets featuring zero net magnetization and non-relativistic spin splitting in the electronic bands \cite{Hayami2019,Smejkal2020,Yuan2020,Mazin2021,Smejkal2022,Smejkal2022_2} provide a natural setting, as they offer a powerful platform for creating topological phases in two and three dimensions \cite{Zhu2023,Heung2024,Ghorashi2024,Hodge2025,Li2025,Chen2025,Feng2025,Antonenko2025}. In the non-interacting case, combining altermagnetism with Ising-type spin-orbit coupling (SOC) is capable of realizing both first- \cite{Zhang2025,Antonenko2025,Chen2025} and higher-order topological phases \cite{Yang2025}, which find applications in spintronics \cite{Bai2023,Ang2023,Sun2023,Chi2024,Zhang2024} and quantum computing \cite{Heung2024,Hodge2025}. 

These topological phenomena are yet to be realized experimentally, despite there being many candidate materials for altermagnets \cite{Krempasky2024,Lee2024,Osumi2024,Mazin2023,Ding2024,Lu2025,Li2025,Ni2010,Fuwa2010,Freelon2019,Wei2025,Jiang2025,Zhang2025}, with key observations including $g$-wave altermagnetic band splitting MnTe \cite{Krempasky2024,Lee2024,Osumi2024} and CrSb \cite{Ding2024,Lu2025,Li2025} in three dimensions and $d$-wave altermagnetic band splitting in the group of the oxycalcogenides in two dimensions \cite{Ni2010,Fuwa2010,Freelon2019,Wei2025,Jiang2025,Zhang2025}. In the latter, recent neutron diffraction experiments observed that certain vanadate compounds in this family exhibit G-type antiferromagnetism \cite{Sun2025_2,Xie2026}, which would lead to spin-degenerate bands. A combined study using neutron diffraction and angle-resolved photoemission spectroscopy (ARPES) studies \cite{Yang2026}, points to an interpretation of these materials as bulk antiferromagnets hosting surface altermagnetism \cite{Lange2026}. The central ingredient for the altermagnetic spin splitting at the surface therein lies in the ``anti-CuO$_2$" structure of the $T_2$O layers, consisting of transition metal ($T$) and oxygen atoms (O) in a Lieb lattice arrangement. Thereby, in the monolayer limit, the altermagnetic spin splitting should persist. For this case, it has recently been theoretically shown that a Lieb lattice Hubbard model for materials with this structure can realize both altermagnetic Mott insulators and metals as their ground state \cite{Kaushal2025}. 

\begin{figure}
    \centering
    \includegraphics[width=0.8\linewidth]{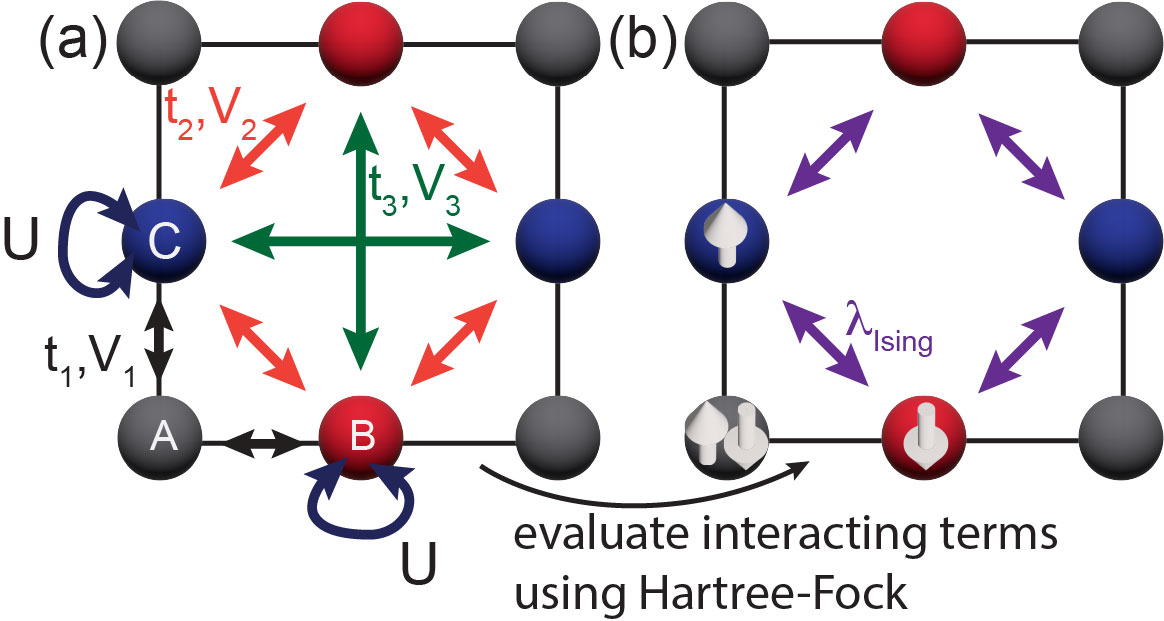}
    \caption{Process for creating a topological altermagnetic correlated insulator from interactions: starting from a (a) Lieb lattice with hoppings and both on-site Hubbard and extended Coulomb interactions, whose interplay generates (b) altermagnetic moments (here shown for average electronic density 4) with an effective Ising-type spin-orbit coupling betweeen the $B$ and $C$ sublattices.}
    \label{fig:Fig1}
\end{figure}

In this Letter, we show that the addition to this model of extended Coulomb interactions naturally yields the emergence of both topological and magetic order, resulting in topological altermagnetic correlated insulating (TACI) states.
In contrast to previous studies, we assume neither the altermagnetic order nor an Ising-type SOC, the crucial ingredient for the topological order, but rather obtain these ingredients self-consistently from the interactions using the unrestricted Hartree-Fock (HF) theory. Thereby, we obtain a simple model for an intrinsic interacting topological altermagnet, paving the way towards finding a suitable material candidate. 
For this purpose, we study the Lieb lattice pictured in Fig.\ \ref{fig:Fig1} with local, repulsive Hubbard interactions present only on the $B$ and $C$ sublattice, which contain the magnetic atoms, while we keep the $A$ sublattice non-magnetic, and also include Coulomb interactions up to those acting between third nearest-neighboring atoms. 

In this setting we use the HF decomposition to map out the topological and magnetic phase diagram as a function of the local Hubbard interaction, the extended Coulomb interaction strengths and the strength of inversion-symmetry breaking Rashba SOC for electron densities of 2 and 4 per unit cell. 
Further, we demonstrate that upon including an SU(2) and inversion symmetry-breaking Rashba term, which chooses a preferred direction for the N\'eel vector, our model allows for the emergence of higher-order topological corner modes purely from interactions.
Finally, we utilize exact diagonalization techniques to confirm the validity of our mean-field theory results.

{\it Model.~} 
We consider a two-dimensional Hubbard model on a Lieb lattice, described by the Hamiltonian
\begin{equation}
\begin{aligned}
    \mathcal{H}_0 =& \;  \varepsilon_A \sum_{i \in A} n_{i} -\mu \sum_{i} n_i + U \sum_{i \in \{B,C\}} n_{i,\uparrow} n_{i,\downarrow}  \\
    &+ t_1 \sum_{\langle i,j \rangle_m, \alpha} c^\dagger_{i,\alpha} c^{}_{j,\alpha} + \sum_{\langle\langle i,j \rangle\rangle, \alpha} t_2 c^\dagger_{i,\alpha} c^{}_{j,\alpha}
    \label{eq:ham_liebHubbard}
\end{aligned}
\; ,
\end{equation}
where $c^\dagger_{i,\alpha}$ creates an electron at site $i$ (which labels both the unit cell and the sublattice) with spin $\alpha$, $n_{i,\alpha} = c^\dagger_{i,\alpha} c^{}_{i,\alpha}$ is the density operator and $n_i = \sum_\alpha n_{i,\alpha}$. $\mu$ is the chemical potential, which we determine such that the desired filling is achieved, $\varepsilon_A$ is a local potential at site $A$.

For this model, it was previously shown \cite{Kaushal2025} using the unrestricted HF theory at filling 2 and 4, that combining nearest and next-nearest neighbor hopping parameters $t_1$ and $t_2$ with the repulsive Hubbard interaction $U$ on sites $B$ and $C$, as depicted schematically in Fig.~\ref{fig:Fig1}(a), leads to antiferromagnetic order on sublattices $B$ and $C$. This in turn gives rise to altermagnetism with an effective time-reversal symmetry $C_{4z} \mathcal{T}$, where $C_{4z}$ denotes the fourfold rotation around the $z$-axis and $\mathcal{T}$ is the usual time-reversal operator. 
\begin{figure}
    \centering
    \includegraphics[width=\linewidth]{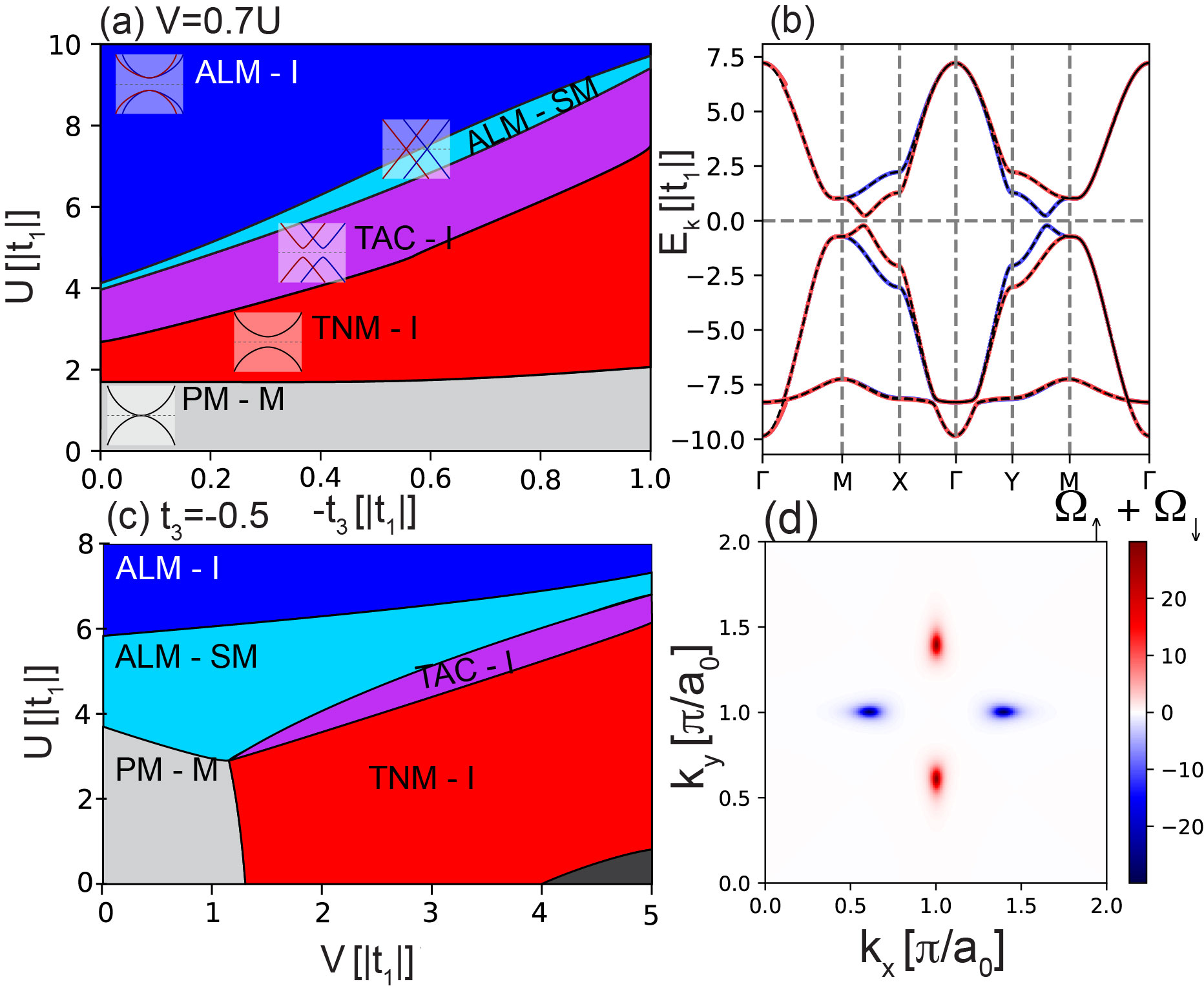}
    \caption{Topological phase diagram (a) as a function of $U (V=0.7U)$ and $t_3$, (b) as a function of $U$ and $V$ for average electronic density 4, (c) band structure in the topological altermagnetic phase with spin-$\uparrow$ ($\downarrow$) polarization indicated in red (blue). (d) spin-resolved Berry curvature in the Brillouin zone. Parameters are $t_1=-1,t_2=1.5,\epsilon_A=0$ with $t_3=-0.5$ in (b), $U,V,t_3=(5.5,3.85,-0.5)$ in (c,d).}
    \label{fig:Fig2}
\end{figure}

{\it Topological Altermagnetic Insulators from Interactions.~} 
On the Lieb lattice, it was previously shown \cite{Antonenko2025} that adding the Ising SOC term $\mathrm{i} \lambda_{I} \sum_{\langle\langle i,j \rangle\rangle,\alpha} (-1)^\alpha  \nu_{ij} c^\dagger_{B,i,\alpha} c_{C,j,\alpha} + {\rm h.c.}$, with $\nu_{ij} = \pm 1$ to an altermagnetic Dirac semimetal opens a gap at the Dirac points. The induced topological phase features a quantum spin Hall effect, described by a spin Chern number \cite{Prodan2009}
\begin{equation}
   C_\sigma = \frac{1}{2\pi \mathrm{i}} \int_{\mathrm{BZ}} \mathrm{d}^2 \mathbf{k} \; \mathrm{Tr} \left\{ P_\sigma (\mathbf{k}) \left[ \partial_{k_x} P_\sigma (\mathbf{k}),  \partial_{k_y} P_\sigma (\mathbf{k}) \right] \right\}  \; ,
\end{equation}
where $P_\sigma(\mathbf{k})$ refers to the projector onto the occupied bands with spin $\sigma$. 

To obtain this state from interactions, we first require an \textit{extended metallic region with altermagnetic correlations}. 
For this purpose, we add hoppings across the Lieb lattice plaquettes between the same sublattice
\begin{equation}
    \mathcal{H}_{t_3} = t_3  \left( \sum_{i \in B ,\alpha} c^\dagger_{i,\alpha} c^{}_{i+\hat{y},\alpha} + \sum_{i \in C,\alpha} c^\dagger_{i,\alpha} c^{}_{i+\hat{x},\alpha} + {\rm h.c.}\right) \; ,
\end{equation}
leading to the presence of Dirac points at the Fermi level. Inducing an Ising SOC in the system then goes hand in hand with a gap emerging at these point. We can obtain this ingredient from a Coulomb interaction $\frac{V}{d_2} \sum_{\langle\langle i,j \rangle\rangle}  n_i n_j$ between $B$ and $C$ sublattices on next-nearest neighbor sites, as under HF decomposition it becomes
\begin{equation}
    \frac{V}{d_2} \sum_{\langle\langle i,j \rangle\rangle} \left(\langle n_i \rangle n_j - \sum_{\sigma, \sigma'} \langle c^\dagger_{i,\sigma} c_{j,\sigma^\prime} \rangle c^\dagger_{j,\sigma^\prime} c_{i,\sigma} \right) \; .
\end{equation}
Here, for an out-of-plane spin orientation, a non-zero imaginary part of the equal-spin Fock term $- \mathrm{i} \frac{V}{d_2} \sum_{\langle\langle i,j \rangle\rangle} \sum_{\sigma} \mathrm{Im}(\langle c^\dagger_{i,\sigma} c_{j,\sigma} \rangle) c^\dagger_{j,\sigma} c_{i,\sigma}$ obeying time-reversal symmetry $\mathrm{Im}(\langle c^\dagger_{i,\uparrow} c_{j,\uparrow} \rangle) = -\mathrm{Im}(\langle c^\dagger_{i,\downarrow} c_{j,\downarrow} \rangle)$ gives exactly the term we need to open a topological gap. 

Note that as our model is SU$(2)$ invariant, there is no preferred spin orientation, and hence the emergent correlations along the bonds between $B$ and $C$ sublattices will reorient together with the orientation of the N\'eel vector.
However, including the Coulomb interaction along this bond by itself favors nematic phases with different occupations of the $B$ and $C$ sublattices, which is detrimental to the altermagnetism. To balance the charge configurations in the unit cell, the minimal model must include the nearest-neighbor and third-neighbor Coulomb interactions as well, so that the overall Hamiltonian becomes 
\begin{equation}
    \mathcal{H} = \mathcal{H}_0 + \mathcal{H}_{t_3} + \sum_{m=1}^3 \frac{V}{d_m} \sum_{\langle i,j \rangle_m}  n_i n_j \; .
\end{equation}
Here $V$ is the strength of the Coulomb interaction, which gets reduced by the distance as $d_1 = 1, d_2 = \sqrt{2}, d_3 = 2$.

We solve the interacting part of the Hamiltonian using the unrestricted HF theory, evaluating the filling of the three sublattices, the spin expectation values on sublattices $B$ and $C$, as well as the Fock terms from the nearest-, next-nearest- and next-next-nearest neighbor correlations self-consistently (for details, see Supplemental Material (SM) Section I).

{\it Topological Phase Diagram for $n=4$.~} 
We use the staggered (total) magnetization
\begin{equation}
    m_{s/t} = |\mathbf{S}_{B} \mp \mathbf{S}_{C}| = \frac{1}{N} \sum_\mathbf{k} |\mathbf{S}_{\mathbf{k},B} \mp \mathbf{S}_{\mathbf{k},C}|
\end{equation}
to quantify the strength of collinear antiferromagnetic (ferromagnetic) order. We start by considering a fixed ratio of the two interaction strengths of $V=0.7U$ and investigate the magnetic and topological phase diagram as a function of $U$ and $t_3$, for which the resulting phase diagram is shown in Fig.~\ref{fig:Fig2}(a), with further details and results for $n=2$  presented in SM Section II.A and II.B. 

For small interaction strengths, we observe a paramagnetic (PM) metal with a quadratic band touching at the $M$ point in the Brillouin zone. As $U$ and $V$ are increased, the extended Coulomb interactions open a gap at the $M$ point, creating a topological non-magnetic (TNM) spin Chern insulator with $C_s = 1$, according to the mechanism proposed in Ref.~\cite{Sun2009}. Upon further increasing both interaction strengths, antiferromagnetic order emerges between the $B$ and $C$ sublattices. This splits the gapped quadratic band closing into two spin-polarized, gapped Dirac points leading to an altermagnetic state with a non-trivial spin Chern number $C_s = 1$, giving a TACI phase. For this phase, we show the HF band structure in Fig.~\ref{fig:Fig2}(c). Here, the spin polarization of the bands is shown in red (blue) for spin $\uparrow$ ($\downarrow$), respectively, indicating that the altermagnetic spin splitting is strongest around the Dirac points, while the bands along the $\Gamma$-$M$ line remain doubly degenerate. The non-trivial spin Chern number in this phase originates from a non-zero spin-resolved Berry curvature, which accumulates around the locations of the gapped Dirac points (see Fig.~\ref{fig:Fig2}(d)). 

Increasing the interaction strengths further, these Dirac points move towards the $X(Y)$ points, which eventually makes the metallic state with gapless Dirac cones energetically preferable, corresponding to an altermagnetic (ALM) semimetal. While this state no longer has a topological gap, the Dirac points have a topological charge of $\pm 1$. Finally, once the Dirac points reach the $X(Y)$ points, a trivial gap emerges, creating an ALM insulator. Because the insulating gap is opened by suppressing double occupancy on the $B$ and $C$ sites, with the charge occupations from the self-consistent calculations very close to 1, we designate this as an altermagnetic Mott insulator, similar to Ref.~\cite{Kaushal2025}.

Next, we consider what happens when we move away from this ratio of $V/U$. For this purpose, we present the phase diagram for $t_3=-0.5$ as a function of $U$ and $V$ in Fig.~\ref{fig:Fig2}(b). Here, we can see that for small $U$, the system becomes either a paramagnetic metal for small $V$ or a TNM insulator for $V$ large enough to induce spin-orbit coupling terms between the $B$ and $C$ sublattice. For large $V$, there exists a third region, where the $B$ and $C$ sublattice are no longer equally occupied and a charge density-wave emerges. Upon increasing $U$, the system enters into the ALM semimetal for small $V$, with the TACI phase emerging between this phase and the TNM phase. Note that this ALM semimetal is unstable towards becoming a TACI, as any small external Ising-type spin-orbit coupling between the $B$ and $C$ sublattice opens a gap at the Dirac points and converts the entire region into a TACI. Finally, for large $U$ the ALM Mott insulator persists and remains topologically trivial even for large $V$, as the insulating gap is too large to be overcome by $V$.  

\begin{figure}
    \centering
    \includegraphics[width=\linewidth]{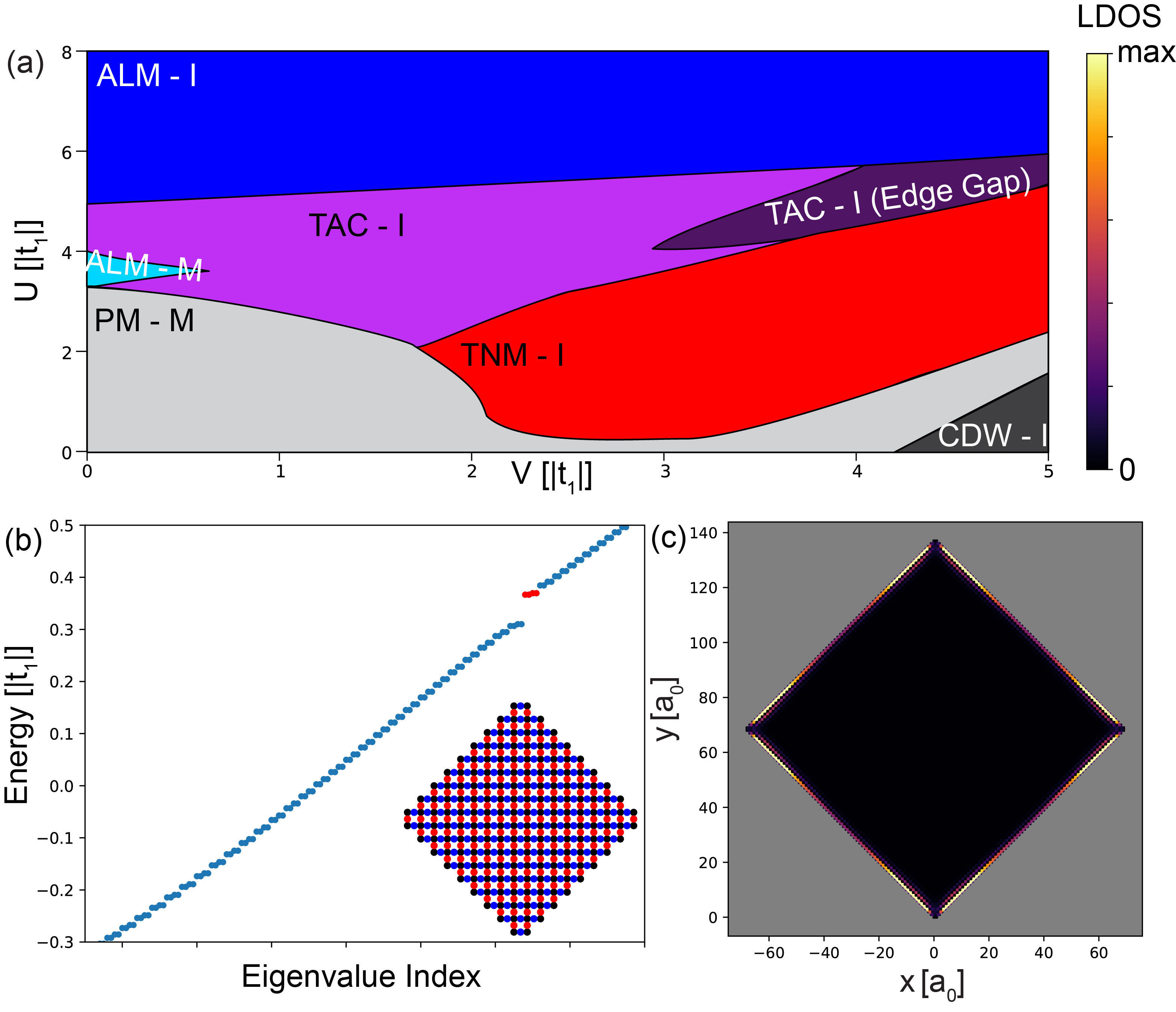}
    \caption{(a) Topological phase diagram for filling $n=4$ with regions hosting corner modes indicated, (b) energy spectrum of a real-space system with diagonal $A$-terminated edges with energies corresponding to corner modes marked in red (see inset), (c) local density of states at the mean energy of the corner modes. Parameters are $t_1=-1,t_2=1.5,t_3=-0.2,\epsilon_A=0,\lambda_1=-0.4,\lambda_3=-0.2$ and $U=5.2,V=4.7$ in (b,c).}
    \label{fig:Fig3}
\end{figure}

{\it Emergence of Corner Modes.~} Similar to our previous study on a crystallographic altermagnet with a non-trivial spin Chern number, the helical edge modes along edges with ferromagnetic structure are separated in momentum, which makes them more robust to inversion-symmetry breaking terms (for further details, see Ref.~\cite{Bedow2026}). However, for the helical edge modes along antiferromagnetic edges, i.e. when an edge is terminated by alternating sublattices or non-magnetic sublattices, Rashba SOC  between nearest-neighbor or third nearest-neighbor atoms on the Lieb lattice of the form 
\begin{equation}
\begin{aligned}
    \mathcal{H}_\lambda &= \mathrm{i}\lambda_1 \sum_{\langle i,j \rangle, \alpha, \beta} ( (\mathbf{d}_{ij} \times\boldsymbol{\sigma})^z_{\alpha\beta}) c^\dagger_{i,\alpha} c^{}_{j,\beta} \\
    &-\; \mathrm{i}\lambda_3  \sum_{\alpha,\beta} \left( \sum_{i \in B} \sigma^x_{\alpha\beta} c^\dagger_{i,\alpha} c^{}_{i+\hat{y},\beta} - \sum_{i \in C} \sigma^y_{\alpha\beta} c^\dagger_{i,\alpha} c^{}_{i+\hat{x},\beta} \right) + {\rm h.c.} \; .
\end{aligned}
\end{equation}
produces an edge gap (see SM Section III for details). 

This hybridization of the edge modes along the diagonal direction can in fact give rise to other interesting physics \cite{Wang2023,Yang2025}, namely when the hybridization gap lies within the insulating gap, giving rise to an edge gap. To demonstrate this, we include both types of Rashba SOC, $\lambda_1$ and $\lambda_3$, and present the topological phase diagram for this case in Fig.~\ref{fig:Fig3}(a). Note that both of these terms preserve the $C_{4z}\mathcal{T}$ symmetry, whereas they break the SU$(2)$ invariance of the magnetic moments, as these now need to rotate together with the lattice under the $C_{4z}$ rotation. As a result, the N\'eel vector acquires a preferred out-of-plane orientation.
Here, we can see that the altermagnetic region with a non-zero spin Chern number now overlaps largely with a region exhibiting a finite gap along diagonal edges. The consequence of this can be observed when considering the system with open boundary conditions and diagonal $A$ terminations in real space, as shown in Fig.~\ref{fig:Fig3}(b). 

The spectrum of this system shows two sets of doubly degenerate states (marked in red) inside the edge gap, which when computing the local density of states shown in Fig.\ \ref{fig:Fig3}(c) correspond to the so-called corner modes \cite{Benalcazar2017}, as they are localized in the corners of the system. These states were recently shown to occur when both altermagnetic order,  Ising and Rashba SOC are present \cite{Yang2025}. Moreover, these corner modes persist for the alternative $BC$-termination (see SM Section III for details). Our results thereby confirm that these two ingredients can co-exist without any assumptions and that the second-order topological state can emerge purely from interactions, while being robust to the termination of the system.

\begin{figure}
    \centering
    \includegraphics[width=0.95\linewidth]{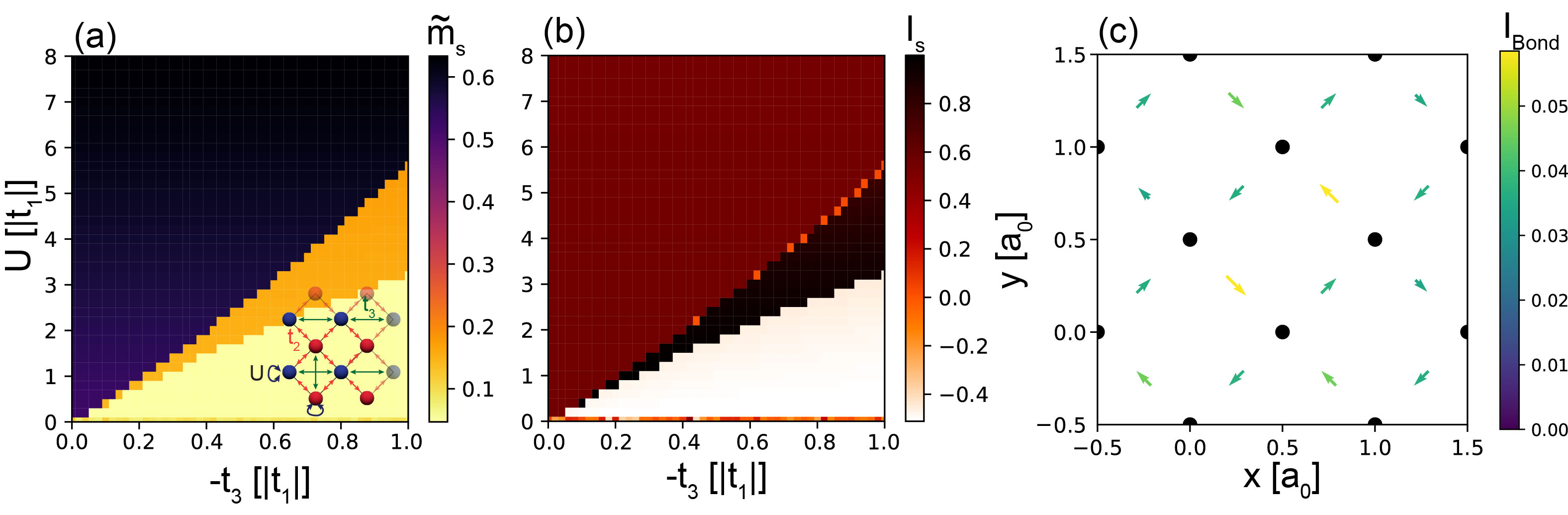}
    \caption{(a) Staggered magnetization, (b) total Ising correlation as a function of $U$ and $t_3$ with $V=0.05U$, (c) bond-wise Ising correlation obtained from exact diagonalization on a $2\times2$ cluster.
    Parameters are $t_2=1.5$ with $U=4.0,t_3=-1.0$ in (c).}
    \label{fig:Fig4}
\end{figure}

{\it Exact Diagonalization.} The band structure in Fig.~\ref{fig:Fig2}(b) for filling $n=4$ shows two bands at the bottom that have a very large spectral weight on the $A$ sublattice (for further details, see SM Section IV.A). This indicates that the $A$ sublattice does not contribute to the formation of the topological phase and motivates a treatment of this system through a checkerboard lattice model, in which we omit the $A$ sublattice (see inset of Fig.~\ref{fig:Fig4}(a) for a schematic and SM Section IV for details of this model). In this case, we can analyze small cluster systems using exact diagonalization to calculate the staggered magnetization from correlations $\tilde{m}_s = \frac{1}{(N_xN_y)^2} \sum_{\alpha \in \{ x,y,z\}} \sum_{i,j} \langle \mathbf{S}_{i,B}^\alpha \left( \mathbf{S}_{j,B}^\alpha - \mathbf{S}_{j,C}^\alpha \right) \rangle$, between $B$ and $C$ sublattices, which lets us identify altermagnetic order. 
To identify the topological character, analogous to Ref.~\cite{Sur2018}, we calculate the current correlations between nearest-neighbor bond $\langle i,j\rangle$ and a reference bond $\langle i_0, j_0 \rangle$ of opposite spin as 
\begin{equation} 
I_{\mathrm{Bond},ij} = - \langle (c^\dagger_{i,\uparrow} c_{j,\uparrow} - c^\dagger_{j,\uparrow} c_{i,\uparrow}) (c^\dagger_{i_0,\downarrow} c_{j_0,\downarrow} - c^\dagger_{j_0,\downarrow} c_{i_0,\downarrow}) \rangle.
\end{equation}
We then consider a staggered sum, where the pre-factors of each bond match the signs the induced Ising spin-orbit coupling $I_s = \sum_{\langle i,j \rangle} \nu_{ij} I_{\mathrm{Bond},ij}$ in the mean-field theory would exhibit.

In Fig.~\ref{fig:Fig4} we present these two quantities for a $2\times2$ checkerboard system, inset Fig.~\ref{fig:Fig4}(a), as a function of $U$ and $t_3$, with $V=0.05U$ and 
the same value for $t_2$ as in Fig.~\ref{fig:Fig2}. Here, we impose periodic boundary conditions only for $t_2$ and $t_3$ hoppings along the $x$-direction to avoid double counting of $t_3$ and $\lambda_3$. We find that there exist 3 principal regions: for small values of $U$ and $V$, the staggered magnetization is almost zero, matching the paramagnetic phase from our mean-field results. 
For high values of $U$, in the top part of the phase diagram, we observe strong staggered magnetization, matching our mean-field results for the altermagnetic Mott insulator. 
In between these two regions, we enter a third region with slightly smaller staggered magnetization. The staggered current correlations exhibit the same three regions, with negative values in the paramagnet, intermediate positive values in the altermagnetic Mott insulator and high positive values in the intermediate region. By looking at the current patterns spatially, which we present for a point in the intermediate phase in Fig.~\ref{fig:Fig4}(c), we find that only the intermediate region shows loop current patterns, as we would expect in the topological altermagnetic phase. For higher values of $U$, there is no loop structure despite there being non-zero correlations, and for smaller values of $U$, the current correlations show significant spatial inhomogeneities (for further details, see SM Sec.~IV). Taken together the loop structure of the current correlations and the presence of staggered magnetization confirm the existence of the TACI phase in exact diagonalization.

{\it Discussion.~}
Our results on the extended Lieb lattice Hubbard model establish the potential for altermagnetism and topological insulating states to emerge simultaneously from interactions. Both Hartree-Fock theory and exact diagonalization support the existence of these phases in a large region of the parameter space. Including Rashba spin-orbit coupling leads to second-order topological states, which recently gained interest in non-interacting studies of altermagnetic spin Chern insulators because of their applications in quantum information theory. Here, they emerge purely from interactions.  Our work provides a starting point for future studies to address whether the proposed higher-order corner modes in superconducting altermagnets \cite{Hodge2025} can be obtained from this class of interacting models. Further, having obtained the Ising-type spin-orbit coupling from Coulomb repulsion naturally leads to the question whether fluctuations in the corresponding order parameter could lead to correlation-driven superconductivity in the altermagnetic semimetal, as this falls into the regime of Gross-Neveu criticality \cite{Stangier2026}.

Our results pave the way for finding altermagnetic topological states from the interplay  between Coulomb repulsion and a realistic lattice geometry. Natural material candidates include  oxycalcogenide compounds whose bulk crystals were recently found to be $G$-type antiferromagnets \cite{Sun2025_2,Xie2026} but could still exhibit altermagnetism in their monolayer form \cite{Yang2026,Lange2026}. Proposals to engineer the Lieb lattice in hybrid structures, either by adsorbing atoms such as fluoride \cite{Ding2026,Cui2026} or placing monolayers onto an antiferromagnetic FeSe substrate \cite{Li2026}, offer additional appealing application of our theory.

{\it Acknowledgments}
This work was supported by the Natural Sciences and Engineering Research Council of Canada (NSERC).

{\bf Data availability}
The data that were obtained in this study are available from the authors on reasonable request. \\

{\bf Competing interests}
The authors declare no competing interests.

\end{document}